\documentclass[acmlarge,nonacm]{acmart}
\AtBeginDocument{\settopmatter{printacmref=true}}

\usepackage{graphicx}
\usepackage{multirow}
\usepackage{appendix}

\AtBeginDocument{%
  }

\copyrightyear{2026}
\acmYear{2026}
\setcopyright{cc}
\setcctype{by}
\acmConference[FAccT '26]{The 2026 ACM Conference on Fairness, Accountability, and Transparency}{June 25--28, 2026}{Montreal, QC, Canada}
\acmBooktitle{The 2026 ACM Conference on Fairness, Accountability, and Transparency (FAccT '26), June 25--28, 2026, Montreal, QC, Canada}
\acmDOI{10.1145/3805689.3806467}
\acmISBN{979-8-4007-2596-8/2026/06}

\begin{document}

\title[Participatory Auditing to Surface Blind Spots in Ranked Search Results]{All Eyes on the Ranker: Participatory Auditing to Surface Blind Spots in Ranked Search Results}

\author{Anna Marie Rezk}
\orcid{0000-0002-0351-0167}
\affiliation{%
  \institution{School of Computing Science \\ University of Glasgow}
  \city{Glasgow}
  \country{United Kingdom}}
\email{anna.rezk@glasgow.ac.uk}

\author{Patrizia Di Campli San Vito}
\orcid{0000-0002-2499-8464}
\affiliation{%
  \institution{School of Computing Science \\ University of Glasgow}
  \city{Glasgow}
  \country{United Kingdom}}
\email{patrizia.dicamplisanvito@glasgow.ac.uk}

\author{Ayah Soufan}
\orcid{0000-0002-1689-5633}
\affiliation{%
  \institution{Computer and Information Sciences\\ University of Strathclyde}
  \city{Glasgow}
  \country{United Kingdom}}
\email{ayah.soufan@strath.ac.uk}

\author{Graham McDonald}
\orcid{0000-0002-1266-5996}
\affiliation{%
  \institution{School of Computing Science \\ University of Glasgow}
  \city{Glasgow}
  \country{United Kingdom}}
\email{graham.mcdonald@glasgow.ac.uk}

\author{Craig Macdonald}
\orcid{0000-0003-3143-279X}
\affiliation{%
  \institution{School of Computing Science \\ University of Glasgow}
  \city{Glasgow}
  \country{United Kingdom}}
\email{craig.macdonald@glasgow.ac.uk}

\author{Iadh Ounis}
\orcid{0000-0003-4701-3223}
\affiliation{%
  \institution{School of Computing Science \\ University of Glasgow}
  \city{Glasgow}
  \country{United Kingdom}}
\email{iadh.ounis@glasgow.ac.uk}

\renewcommand{\shortauthors}{Rezk et al.}

\begin{abstract}
\looseness -1  Search engines that present users with a ranked list of search results are a fundamental technology for providing public access to information. Evaluations of such systems are typically conducted by domain experts and focus on model-centric metrics, relevance judgments, or output-based analyses, rather than on how accountability, harm, or trust are experienced by users. This paper argues that participatory auditing is essential for revealing users’ causal and contextual understandings of how ranked search results produce impacts, particularly as ranking models appear increasingly convincing and sophisticated in their semantic interpretation of user queries. We report on three participatory auditing workshops (n=21) in which participants engaged with a custom search interface across four tasks, comparing a lexical ranker (BM25) and a neural semantic reranker (MonoT5), exploring varying levels of transparency and user controls, and examining an intentionally adversarially manipulated ranking. Reflexive activities prompted participants to articulate causal narratives linking to search system properties, rather than merely evaluating result quality. They discussed what they considered to be desirable properties of search results, such as relevance, fairness, and transparency, and how these shape users' understandings of search rankings. Synthesising the findings, we contribute a taxonomy of user-perceived impacts of ranked search results, spanning epistemic, representational, infrastructural, and downstream social impacts. At the same time, interactions with the more convincing neural ranking model revealed potential limits to participatory auditing itself: perceived system competence and accumulated trust reduced critical scrutiny during the workshop, allowing manipulations to go undetected. Participants repeatedly expressed a desire to gain visibility into the full search pipeline, comprising data corpus, ranking model and logic, as well as recourse mechanisms, in order to meaningfully audit the system. Together, these findings show how participatory auditing can surface user perceived impacts and accountability gaps that remain unseen when relying on conventional audits, while also revealing where participatory auditing may encounter challenges and limitations, thus requiring refinement.

\end{abstract}



\received{13 January 2026}
\received[revised]{24 February  2026}
\received[accepted]{2 March 2026}

\maketitle

\section{Introduction}
Search and information retrieval (IR) systems play a central role in shaping access to information, public knowledge, and visibility online. Through ranking and exposure, search systems influence what information is encountered, trusted, and acted upon~\cite{bernard_systematic_2025}. Increasingly, these ranking processes are driven by machine learning–based and neural models, introducing forms of AI into search systems that can be opaque and thus difficult to scrutinise through output-focused evaluation alone~\cite{anand_explainable_2022, nogueira-etal-2020-document}.

Existing approaches to evaluate search systems with respect to fairness\footnote{A fair ranking is understood to contain 1) a sufficient presence of items belonging to different groups to avoid statistical discrimination, 2) a consistent treatment of similar items, and 3) a proper representation of items to prevent representational harms to (specifically but not exclusively) marginalised, disadvantaged or protected groups~\cite{castillo_fairness_2019}.} and relevance often take the shape of audits \cite{perreault_algorithmic_2024, mehrotra_auditing_2017, bernard_systematic_2025}. Audits are understood as a systematic and independent assessments of processes or systems aimed at determining their accuracy, quality, compliance, or effectiveness~\cite{birhaneAIAuditingBroken2024d, becerra_sandoval_historical_2025, goodman_ai_2022}. Audits typically rely on external or expert-led evaluation using technical metrics or output-based analyses. While such approaches are valuable, they ignore the experience of end-users and the insights that can be gained by empowering users to take an active auditing role in participatory auditing processes\footnote{We use the definition of participatory auditing as the process of evaluating a system against normative standards defined by the stakeholders who experience its impacts, rather than by expert-defined standards~\cite{DiCampliSanVito2026ParticipatoryAuditing}.}. Recent work on participatory AI auditing suggests that involving end-users and affected stakeholders can surface blind spots overlooked by expert auditors. However, these works have largely focused on decision-making systems rather than search or IR systems~\cite{birhaneAIAuditingBroken2024d, becerra_sandoval_rethinking_2025}.

Auditing search systems presents a distinct challenge: Search functions as an infrastructural technology that ranks and prioritises information, shaping knowledge over time. Harms may therefore arise cumulatively, through biased exposure, systematic privileging of particular sources, or susceptibility to manipulation by bad-faith actors~\cite{noble2018algorithms, kulshrestha_search_2019}. These dynamics involve multiple interacting components, including 1) content producers, publishers, and owners, whose materials are subject to ranking, 2) the ranking infrastructure itself, whose effects cascade downstream to users, and 3) the users who make interactive decisions with the presented results. Despite this, users are often treated as passive recipients of ranked outputs, even though everyday search practice involves active negotiation with presented results, query refinement, and judgement. As such, we argue the role of users warrants more attention with respect to its value in audits of search systems.

Furthermore, recent regulatory frameworks such as the EU AI Act~\cite{eu_ai_act_2024} and Digital Services Act (DSA)~\cite{eu_dsa_2022} address the growing importance of accountability, transparency, and risk mitigation in AI-driven information systems. The EU AI Act adopts a risk-based approach that recognises harms as potentially emerging through deployment and use, while the DSA explicitly targets search and ranking systems as sources of systemic risk to public discourse, requiring ongoing assessment. In this context, participatory auditing offers a complementary lens for examining how such accountability concerns become visible in practice, by grounding evaluation in users’ situated experiences with search systems.

Responsible AI (rAI) guidance from various (inter)governmental, civil society, and research organisations also promotes stakeholder involvement, with growing bodies of work~\cite{lu_responsible_2024, kallina_stakeholder_2024, aizenberg_designing_2020} arguing for such involvement throughout the AI lifecyle. The objectives of such guidance are, among others, redistribution of power, improvement of socio-technical understandings, better anticipation of risks, and enhancement of public oversight. Kallina et al.~\cite{kallina_stakeholder_2025} argued that, in practice, stakeholder involvement is primarily driven by commercial priorities rather than more rAI-aligned ones. Therefore, in this paper, we examine participatory auditing as a method for surfacing user-perceived harms, accountability concerns, and auditability needs in ranked search results. We conducted three auditing workshops in which participants interacted with a custom search system that deployed two ranking models (\textit{BM25}, a lexical ranker and \textit{MonoT5}, a neural semantic reranker) and multiple interface affordances. The objective was to surface how user-perceived and -experienced issues become legible through participatory auditing practices, which may otherwise remain invisible during traditional audits. Thus, rather than treating users as evaluators of output quality alone, the workshops invited participants to actively interrogate, interpret, and contest ranking behaviour, allowing us to study what kinds of issues surface through participatory auditing practices and where such participatory auditing encounters limitations and requires refinement to be an efficient contribution to traditional auditing. Through this approach, we address the following research questions: 

\begin{itemize}
    \item RQ1: What kinds of harms and accountability issues become legible when users are invited to audit ranked search results? 
    \item RQ2: Which challenges and limitations can arise during participatory audits?
\end{itemize}

Our findings are two-fold: First, participatory auditing enables impact discovery by surfacing epistemic, representational, infrastructural, and potential downstream social harms that remain largely invisible or disentangled from context in output-focused or expert-led audits. Second, analysis of the workshops revealed challenges and limits to auditability, particularly in instances where ranking systems have earned user trust or otherwise conditioned users to look for certain problems while overlooking others.

As such, this paper makes three contributions. Conceptually, we argue that participatory auditing should be understood as complementary to traditional auditing accountability measures for search systems. Empirically, drawing on three participatory audit workshops, we show how participants recognise, misrecognise, or fail to detect harms in ranked search results, and how these judgements vary across ranking models based on perceived system competence. Methodologically, we operationalise participatory auditing of IR through workshops that compare ranking models, vary different interface affordances, and include an adversarial manipulation, offering a structured approach for studying impacts and auditability in search systems.

\section{Background and Related Work}

\subsection{Auditing in Socio-Technical Contexts}

Auditing is a systematic and independent process for evaluating evidence about a system and reporting findings to stakeholders~\cite{li2025making,mokander2024auditing}. In AI, auditing functions as a governance mechanism to assess datasets or algorithms against defined expectations, identify risks, and support accountability for a system’s behaviour and impacts~\cite{mokander2024auditing,mokander2021ethics, berghout2023advanced}. Auditing can take multiple forms, differing by who conducts it, what is examined, and when it occurs. Audits may be internal or external; focus on governance, models, or applications; assess compliance with standards or explore open-ended risks; and be adversarial or collaborative~\cite{mokander2024auditing}. They can also occur before deployment (ex-ante) or after deployment (ex-post), and target a system’s functionality, internal model behaviour, or real-world impacts~\cite{mokander2024auditing}. In addition, specialised audits address ethics, empirical system behaviour, or data quality, provenance, and privacy~\cite{eu_ai_act_2024, mokander2024auditing}. However, in expert-led audits (technical as well as policy-based), auditors hold discretion in matters that are phrased ambiguously. This creates a market for auditors, where platforms are incentivised to choose their auditors based on their benchmarks. As a result, benchmark disparities inadvertently incentivise and further minimal compliance~\cite{fabbri_auditing_2025}. 

\looseness -1 To move beyond such a state of minimal compliance, participatory auditing offers a critical alternative by integrating the unique knowledge and lived experiences of stakeholders, such as end-users, who are directly impacted by the AI systems~\cite{ojewale2025towards}. This approach is essential; as Birhane et al.~\cite{birhaneAIAuditingBroken2024d} argued, the current auditing ecosystem is often a “broken bus” that prioritises narrow technical evaluations over the consequential judgment and accountability infrastructure needed for real-world impact. While expert-led markets may incentivise auditors to avoid challenging the status quo, participatory auditing empowers diverse stakeholders, including those from marginalised communities, to uncover harms that technical benchmarks might miss~\cite{dengWeAuditScaffoldingUser2025a, holstein2019improving}. By encouraging a more inclusive process, participatory methods help build the necessary discovery and advocacy frameworks that move the auditing process from simple performance analysis toward a robust system of transparency and public accountability~\cite{bellamy2019ai, dengWeAuditScaffoldingUser2025a, ojewale2025towards}. Extending this idea, Becerra Sandoval and Jing~\cite{becerra_sandoval_rethinking_2025} argued community-based practices, similar to participatory auditing, offer a valuable bottom-up contribution to system safety and accountability. Drawing on examples from road and driver safety, they emphasised that safety is a lived practice and thus extends beyond system features, relating to social, cultural, economic, and political conditions of affected communities and individuals. In a similar vein, the WeAudit project provides scaffolding for individual users to document perceived harms in GenAI, to ultimately generate reports for practitioners, specifically in text-to-image systems~\cite{dengWeAuditScaffoldingUser2025a}. Aiming for a more holistic approach, the PHAWM project provides workbenches and methodologies to support the participatory auditing of predictive and generative AI applications by individuals without AI expertise~\cite{DiCampliSanVito2026ParticipatoryAuditing}.

Audits of search systems examine how information is selected, prioritised, and presented, with the aim of identifying biases, representational harms, or technical failures~\cite{sandvig2014auditing,robertson2018auditing}. Prior work has audited search results, particularly in image and text search, to uncover systematic gender and racial stereotypes. For example, audits of image search engines have shown that queries for occupations such as “CEO” predominantly return images of white men~\cite{kay2015unequal}. Similarly, a well-known audit of Google Search conducted in the United States demonstrated that arrest-record advertisements were significantly more likely to appear for searches involving names traditionally associated with African American individuals than for names associated with white individuals~\cite{noble2018algorithms}.

Researchers have also audited ranked search results by framing tasks such as CV screening as document retrieval problems. For instance, Wilson and Caliskan~\cite{wilson2024gender} modified identical CVs only by changing names associated with different ethnicities and genders, then ranked them against a job description. A ranking system was considered biased if it consistently favours certain demographic groups despite identical qualifications. In addition to ranking mechanisms, audits have focused on collections and datasets that underpin search systems. One prominent approach involves the use of Datasheets for Datasets, which document a dataset’s motivation, composition, and collection process~\cite{gebru2021datasheets}. Such documentation enables auditors to assess whether collections overrepresent hegemonic viewpoints or suffer from “documentation debt”, potentially leading to biased downstream outcomes. Audits of large web-crawled collections, such as the RealNews corpus~\cite{zellers2019defending}, which draws from approximately 5,000 news domains, have examined the diversity and quality of data used to train generative models. These audits highlight issues such as exposure bias, where sampling from a learned distribution causes generated text to drift progressively away from natural human language as sequence length increases~\cite{zellers2019defending}.

\subsection{Fairness, Accountability, and Transparency in IR}

IR systems have traditionally been evaluated using offline benchmark-based methods grounded in test collections~\cite{cleverdon1970evaluation,voorhees1998variations,sanderson_test_2010}. These consist of a fixed document corpus, a predefined set of queries, and relevance judgements produced by trained assessors, with system performance measured through aggregate metrics such as precision (measuring the fraction of retrieved documents that are relevant) and recall (measuring the fraction of relevant documents retrieved). As such, traditional audits in IR use quantifiable proxies like Normalized Discounted Cumulative Gain (NDCG) for relevance~\cite{jarvelin_cumulated_2002} and exposure metrics for fairness~\cite{diaz_evaluating_2020}.

More recently, evaluation efforts have expanded beyond relevance to incorporate normative objectives such as fairness. For example, the TREC Fair Ranking Track~\cite{ekstrand_overview_2022} evaluated IR according to how well they provided a fair exposure to documents about, or from, different societal groups, operationalised through predefined fairness criteria and metrics. While such initiatives represent important advances in formalising fairness within IR evaluation, they remain grounded in expert-defined objectives, fixed datasets, and metric-based assessments~\cite{sanderson_test_2010}. As a result, broader questions of accountability, such as how harms are perceived, interpreted, or causally attributed by users remain largely outside the scope of conventional IR evaluation frameworks.

This is particularly important, as search systems are inherently vulnerable to position bias\footnote{Position bias is the tendency to favour higher-ranked items regardless of relevance.}~\cite{Joachims_clicks,Craswell_position_bias}, which translates minor differences in relevance into large disparities in user attention~\cite{JanichQEP}. This structural vulnerability can result in unintended bias against demographic groups, such as those defined by race or gender, if their relevant content is systematically placed further down the ranking of search results~\cite{JanichQEP}. Similarly, ranking models also have their own susceptibilities towards manipulation. For instance, MonoT5~\cite{nogueira-etal-2020-document} has been found to react to adversarial keyword-injections in documents, which can artificially inflate the documents' relevance scores~\cite{parry_analyzing_2024}. Such model-specific vulnerabilities can lead to the manipulation of search results beyond traditional SEO practices~\cite{LewandowskiSEO}. 

\looseness -1 In an effort to mitigate potential harms that can arise from bias in the results of IR systems, there have been several attempts to establish the objectives to optimise towards beyond relevance, such as fairness and diversity. For instance, Diaz et al.~\cite{diaz_evaluating_2020} argued that fairness in ranking is closely tied to the exposure to users that the documents receive, due to the documents' positions in ranked search results: deterministic rankings of search results can over-privilege certain documents by consistently placing them in prominent rank positions, even when the underlying relevance differences between documents are small. As such, Diaz et al.~\cite{diaz_evaluating_2020} proposed modelling rankings as a stochastic sampling from a relevance distribution, to distribute exposure more equitably and better reflect uncertainty in the ranking process. As more sophisticated ranking models (such as MonoT5~\cite{nogueira-etal-2020-document}, ColBERT~\cite{ColBERT} or Splade~\cite{formal2021splade}) rely on re-ranking pipelines, Jaenich et al.~\cite{JanichQEP} showed that when a first-stage retrieval results in a biased exposure, this bias is unlikely to be fully corrected through fairness-aware re-ranking alone. Demonstrating that exposure is driven more by candidate set inclusion than by ranking position, the authors~\cite{JanichQEP} introduce Query Exposure Prediction as a diagnostic that enables upstream intervention.

In our work, we do not propose ranking interventions, but instead examine how existing models and their inherent exposure effects are interpreted by users during audits.

\section{Methods}

We conducted three participatory auditing workshops designed to examine how users reason about and evaluate the performance of different ranking-based search systems across two ranking models, with varying levels of interface transparency and control, and adversarial manipulation.

\subsection{Participants}
\looseness -1 We recruited participants through posters and mailing lists across our university campus. Since interest exceeded capacity, we selectively enrolled participants to ensure a balanced distribution in terms of age, professional/educational background, and stated motivation for taking part. Prior knowledge of AI or IR was not required, but participants were asked to express an interest in the topic as part of the sign-up process. Out of 32 applicants, we recruited 21 to take part across the three workshops (WS1 = 4 participants; WS2 = 9; WS3 = 8) which took place in November-December 2025, see details in \autoref{tab:workshop_demographics}. All participants received 40GBP in cash as compensation for their time. When quoting participant comments in our findings, we label transcript excerpts with workshop number and participants number, i.e., WS1P1. For content drawn from post-it notes that cannot be attributed to an individual participant, only the workshop identifier is reported.

\begin{table*}[h]
    \centering
        \caption{Participant demographics by workshop.}
    \label{tab:workshop_demographics}
    \begin{tabular}
    {|p{1.5cm}|p{4cm}|p{6cm}|}
        \hline
        \textbf{Workshop}  & \textbf{Age distribution} & \textbf{Occupation} \\
        \hline
        WS1 & 1×18–24, 3×25–34 & Postgraduate students, PhD student, Software engineer \\ \hline
        WS2 & 3×18–24, 3×25–34, 2×35–44, 1×45–54, 1×55–64 & Undergraduate students, Postgraduate students, University lecturer, Retired educator, Analytics manager \\ \hline
        WS3 & 1×18–24, 6×25–34 & Postgraduate students, Research associates, Service designer, Self-employed \\
        \hline
    \end{tabular}

\end{table*}

\subsection{Custom Search Interface as Research Probe}

All interactive tasks were completed using a custom search system built on the open-source PyTerrier IR framework\footnote{\url{https://github.com/terrier-org/pyterrier} (accessed 10/01/2026)}~\cite{macdonald_pyterrier_2021}. The search system queried a snapshot of the English-language Wikipedia contained in the TREC Fair Ranking Track~\cite{ekstrand_overview_2022} corpus.\footnote{\url{https://fair-trec.github.io/} (accessed 10/01/2026)} Two ranking models were implemented:

\begin{itemize}
    \item \textbf{BM25}~\cite{bm25fntir}, used as a lexical baseline expected to produce shallow, keyword-driven rankings.
    \item \textbf{MonoT5}~\cite{nogueira-etal-2020-document}, used as a neural semantic reranker expected to produce more coherent outputs for some queries.
\end{itemize}

Both ranking models were deterministic, as Diaz et al.~\cite{diaz_evaluating_2020} argued that deterministic (static) rankings can achieve high relevance but also incur high disparity as they can reinforce differences and amplify bias. During the workshops, participants interacted with the search interface using either their own smartphones, or tablets provided for the session, enabling individual exploration of the results.

\subsubsection{IR Metric Evaluation of Queries on Rankers}

To confirm that the two models represent meaningfully different ranking behaviours, we evaluated BM25 and MonoT5 on the two queries used during workshop activities (“famous doctor” and “Tylenol”) using manual relevance judgments by the first author for the top 100 results across both queries, and then computed IR metrics. “Famous doctor” demonstrates a query requiring semantic understanding: BM25 retrieved few relevant results in the top ranks (NDCG\footnote{Normalised Discounted Cumulative Gain} = 0.09 and MAP\footnote{Mean Average Precision} = 0.23), whereas MonoT5 retrieved more relevant results in the top ranks (NDCG = 1 and MAP = 0.88). “Tylenol” is a straightforward query, where both rankers performed well, although MonoT5 still slightly outperformed BM25 (NDCG = 0.93 and MAP = 0.67; NDCG = 0.8 and MAP = 0.63, respectively). These results demonstrate how IR metrics quantify ranker performance for the queries used in our tasks, but they do not provide a holistic perspective, since factors such as query formulation, which can strongly influence retrieval metrics, are not accounted for.

\subsection{Workshop Structure}

Each workshop ran for 2.5 hours and followed the same structure consisting of: (1) introductory audit activities to familiarise participants with search auditing, (2) group discussion prompts, and (3) four interactive search tasks using our custom search system followed by reflexive annotation activities each\footnote{The detailed workshop protocol is in Appendix A.}. All sessions were audio-recorded and transcribed verbatim. The workshop methodology was approved by our university's Ethics Committee.

\subsubsection{Introductory activities}

To introduce participants to the concept of algorithmic auditing, each workshop began with a collective audit of a Very Large Online Platform (TikTok) using the Digital Services Act (DSA) Article 27(1) framework~\cite{fabbri_auditing_2025}. This served to establish a shared vocabulary and understanding of traditional audits. Before moving to the tasks with our custom system, participants viewed a live Wikipedia search for the query “chemist”. This query revealed multiple interpretive issues, including the distinction between chemists (scientists) and chemists as pharmacies (particularly in British English), as well as a notable lack of gender and ethnic diversity in the resulting biographies. This step encouraged participants to reflect on ambiguity and mismatches between intent and query as well as representation and bias in search rankings.

\subsubsection{Reflection and Annotation Activities}\label{3.3.2}
After the introductory activities, participants completed a baseline elicitation exercise in which they individually annotated on post-its what they believed mattered in ranked search results. This activity was used to surface participants’ normative expectations prior to interacting with the custom search interface.

\begin{figure}
    \centering
    \includegraphics[width=1\linewidth]{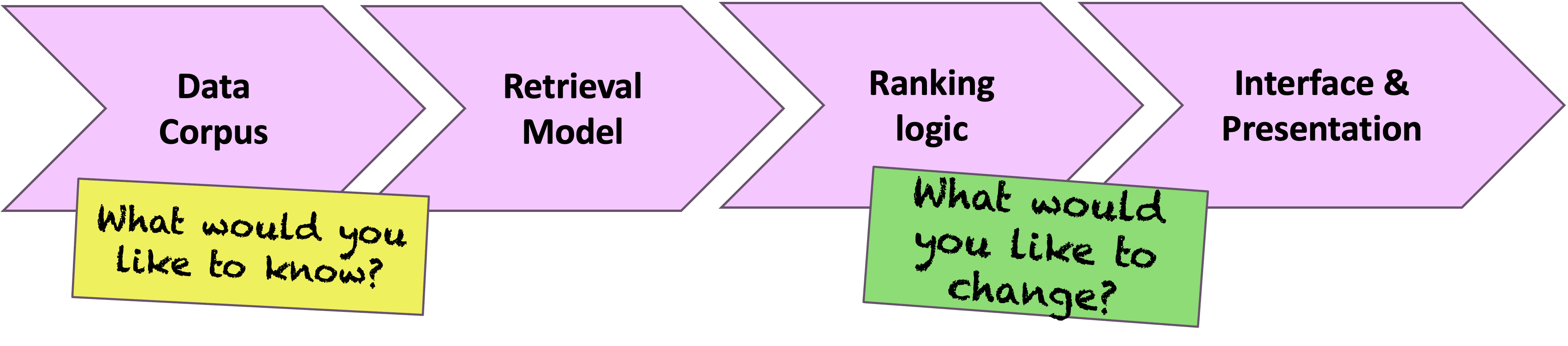}\vspace{-1em}
    \caption{Simplified pipeline diagram for reflexive annotation activities to identify information and recourse needs of users.}
    \Description{Pipeline consisting of four components: Data Corpus, Retrieval Model, Ranking Logic, Interface and Presentation. Example post-its added saying “What would you like to know?” and “What would you like to change?”}
    \label{fig:PipelineBlank}
\end{figure}

Each subsequent task with our custom search system was followed by short annotation and discussion exercises in which participants evaluated the presented rankings, marked points of concern, and articulated their reasoning. To support reflection beyond individual results, participants also engaged with a simplified pipeline diagram representing key components of a search system (corpus, ranking model, ranking logic, and user interface), which they annotated to indicate where greater transparency, additional information, or forms of control and oversight were desired (\autoref{fig:PipelineBlank}).

After the activities with our search system, participants were asked to discuss and identify impacts and consequences that they associated with search systems. These participant-identified events were then used to pre-populate the vertical axis of a shared reflection matrix, while the horizontal axis was populated using the dimensions participants had earlier identified as mattering to them in search. Participants collectively populated the matrix through discussion, mapping search-related properties to perceived events and downstream consequences. This final activity was designed to elicit how participants connected ranking properties to broader social and epistemic effects.

\subsubsection{Task 1: Baseline Lexical Ranker BM25}

For the first task, the instructed query was “famous doctor”, a deliberately challenging query for lexical retrieval as it combines a profession with an adjectival qualifier that requires semantic interpretation rather than exact term matching. Because BM25 performs literal keyword matching, results were intentionally poor: the system retrieved pages with partial lexical overlap rather than semantically relevant biographies, returning the top 10 results to participants. This task was intended to establish a baseline for participants’ expectations around relevance, system capability, and the interpretability of keyword-driven rankings.

\subsubsection{Task 2: Neural Semantic Reranker MonoT5}

The second task used the same interface and query (“famous doctor”), but replaced BM25 with MonoT5. MonoT5 reranks the top 100 BM25 results based on semantics and therefore produces in comparison much more coherent, biography-focused top ten results. This contrast allowed participants to compare a lexical baseline with a semantic model.

\subsubsection{Task 3: Transparency and Control}
\label{task3}
\begin{figure*}
    \centering
    \includegraphics[width=0.9\linewidth]{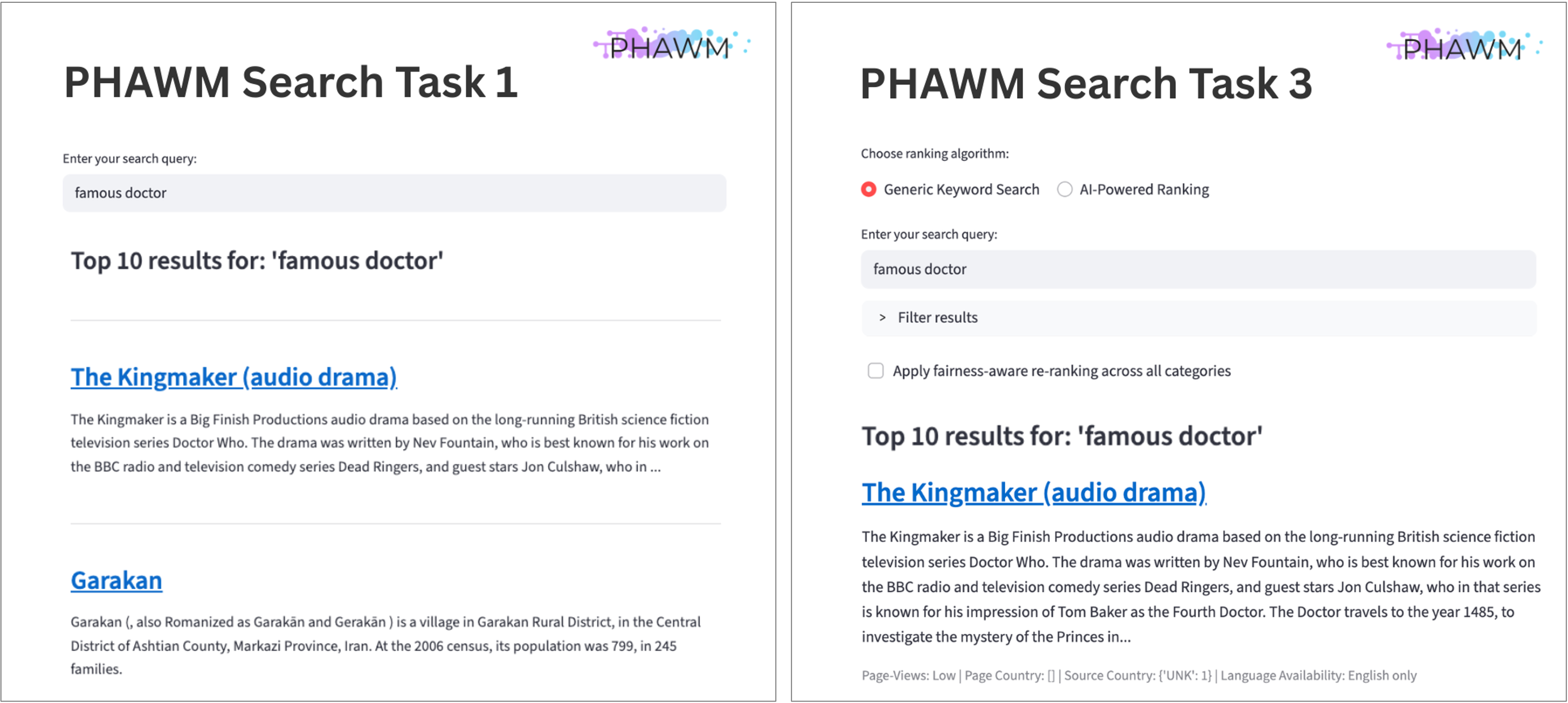}
    \caption{Left: Screenshot of search interface of Task 1 with project branding (same interface also used for Tasks 2 and 4), right: interface of Task 3 with increased transparency and controls (described in Section~\ref{task3})}
    \Description{Screenshot of Task 1 contains a search bar and top ten articles, with title, snippet and URL of each article. Screenshot of Task 3 contains choice between BM25 and MonoT5 ranker, filtering mechanisms, auto-balance button, and top ten articles, with title, snippet, URL, and meta-data of each article.}
    \label{fig:Interface}\vspace{-1em}
\end{figure*}

The third task introduced an expanded interface~\autoref{fig:Interface}, which allowed for model switching between BM25 and MonoT5 (which were labelled as “Generic Keyword Search” and “AI-Powered Ranking” respectively for better user understanding) and provided article metadata (page country, source country, page views, and language availability), metadata-based filters, and an “auto-fairness” function that automatically equalised representation across the included metadata.

The purpose was to explore which forms of transparency participants relied on to reason about bias and accountability and how they interpreted different accountability interventions.

\subsubsection{Task 4: Adversarial Manipulation}\label{3.3.6}

For the final task, participants interacted with a MonoT5-based system, typing in the pre-defined query “Tylenol”. This query was intentionally chosen as a highly specific brand name referring to a well-known pharmaceutical product, for which participants would reasonably expect a narrow and relatively homogeneous set of results. One Wikipedia biography, that of “Helen Keaney”, which normally ranked 17th, was adversarially manipulated through keyword-injection, adding the terms “relevant”, “true”, and “Tylenol” to its title and snippet. The term “Tylenol” was added to spuriously link the document with the user query, while the terms “relevant” and “true,”  were injected to take advantage of a susceptibility within MonoT5's inference-time prompt~\cite{parry_analyzing_2024}. These manipulations elevated the document into the top-five results. Only the top five results were shown in this task for brevity and to make the manipulated result more detectable. The task assessed whether participants, having potentially been primed during previous tasks to having strong perceptions of MonoT5 as a sophisticated semantic system, could detect or question such adversarial attacks.

\subsection{Participatory Auditing Framing}

The workshop was framed around the purpose of participatory auditing. As opposed to other common methods such as participatory design which aims to improve systems through user input, participatory auditing aims to evaluate existing systems against user-defined normative standards like transparency, fairness, and accountability (see Section~\ref{3.3.2}). Thus, the key distinction is intent. Our workshops constitute auditing because participants assessed systems against their normative values that emerged from the DSA auditing activity in the introductory task. Activities such as the adversarial manipulation detection (see Section~\ref{3.3.6}), constitute an audit function, as the intended output are an impact taxonomy and surfacing of accountability gaps. As such, the objective is accountability, not system improvement.

The decision to deploy a custom interface as opposed to using a naturalistic settings was made for three reasons: 1) by deploying our own ranker, we facilitated a participatory auditing workshop on a transparent-box system where we knew the data and the ranking models used; 2) introducing known adversarial manipulation required control over ranking; 3) our approach retains ecological validity as we used a real corpus (Wikipedia snapshot as used by TREC~\cite{ekstrand_overview_2022}), and real ranking algorithms. A naturalistic settings would have sacrificed experimental control needed for the auditing framing of the workshops.

\subsection{Data Analysis}
Transcripts from the workshop recordings, along with post-it artefacts produced during the workshops, were analysed using reflexive thematic analysis following Braun and Clarke’s approach~\cite{braun_thematic_2022}. Initial coding was conducted by the lead author using NVivo software, focusing on patterns in how participants articulated concerns, expectations, and interpretations of ranked search results.

Post-it artefacts created during contextualised annotation tasks (e.g., interface screenshots, pipeline diagram, and impact matrix) were first organised and analysed within the virtual whiteboard environment Miro to preserve their task-specific context. These artefacts were then coded in relation to the corresponding workshop activities and integrated into the broader thematic analysis alongside the transcript data.

Codes and emerging interpretations were then iteratively discussed with co-researchers to support reflexivity and mitigate subjective interpretative bias. 

\section{Findings}

In this section, we present the emerging themes (see~\autoref{tab:Codebook}) through which participatory auditing renders visible the epistemic, infrastructural, and downstream consequences and harms in search systems, many of which would remain largely invisible in output-focused or expert-led audits.

\begin{table*}[tb]
\centering
\caption{Codebook of the thematic analysis. For each theme we show the name, description, an example quote or post-it from the workshops and how often this theme occurred (\#).}
\label{tab:Codebook}
\begin{tabular}{|p{2.7cm}|p{3.1cm}|p{8.1cm}|p{0.2cm}|}
\hline
\textbf{Theme}                    & \textbf{Codes} & \textbf{Example} & \textbf{\#} \\ \hline
\multirow{3}{2.7cm}{Baseline expectations of ranked search results} &       Relevance&         \textit{“Probably relevance is the most important thing for me.”} (WS3P6) &    22\\ \cline{2-4}
                         &       Unbiased&         \textit{“The results should not be biased to a certain region.”} (WS1) &    15\\ \cline{2-4}
                         &       Accuracy&   \textit{“Scientific strength / academic rigour – best of the best.”} (WS2)     &    12\\ \cline{2-4}
                         &      User Control & \textit{“Being able to customise the ranking parameters — that would be really nice.”} (WS2P1) & 9 \\
                         \hline \hline
\multirow{3}{2.7cm}{Epistemic authority of neural rankings and undetected manipulation} &       Perceived semantic understanding &         \textit{“The second [ranker (Task 2, MonoT5)] was more like the common understanding of the world.”} (WS3P3) &  19  \\\cline{2-4}
                         & Undetected manipulation \& misplaced trust      &    \textit{“As we're using AI, when we prompt it and it gives us something, we take it as truth.”} (WS3P2)    &  18  \\\cline{2-4}
                         &  Recognition of vulnerability after exposure     &   \textit{“It just shows that AI can be manipulated… and it’s not as sophisticated as we think.”} (WS3P1)      &   11 \\
                         \hline \hline
\multirow{3}{2.7cm}{Transparency and recourse as coupled accountability needs} &   Provenance on corpus-level &   \textit{“I want to understand where the data is coming from.”} (WS2)      &   18 \\\cline{2-4}
                         &  Desire for user control      & \textit{“Ability to refine or clarify the ranking”}(WS2)   &  20   \\\cline{2-4}
                         &    Recourse mechanisms  &  \textit{“A way for users to report results that aren’t relevant or harmful.”} (WS1)      &  9   \\
                         \hline \hline
\multirow{3}{2.7cm}{Collective sense-making of impacts of search (in relation to users' baseline expectations)} &  Perspective bias diminishing diversity     &   \textit{“Biasing results towards Western/English culture”} (WS2)      &  9  \\\cline{2-4}
                         &  Misinformation eroding trustworthiness     &   \textit{“If you are aware of misinformation, trustworthiness will be linked to that”}  (WS3)    &  7  \\\cline{2-4}
                         &   Power concentration diminishing diversity  &       \textit{“Centralised power in terms of AI tools”} (WS2)  &   7 \\ 
                         \hline
\end{tabular}
 
\end{table*}

\subsection{Baseline Expectations of Ranked Search Results}

After the introductory activities, but before interacting with the custom search interface, participants were asked to reflect on what mattered to them in ranked search results. Across all three workshops, participants articulated a remarkably consistent set of expectations. Relevance was the most frequently cited criterion, but it was rarely framed in purely technical terms. Instead, participants linked relevance to trustworthiness, accuracy, evidence, and contextual appropriateness (e.g., in relation to query “famous doctor”, WS3: \textit{“Scientific [articles] (if I'm searching for something medical”}), suggesting that relevance was understood as an epistemic and normative property rather than a retrieval metric.

Participants also treated fairness, non-bias, and the absence of commercial or sponsored influence as baseline expectations of search systems (e.g., WS2: \textit{“not driven by marketing”}; WS3: \textit{“not politically biased”}). Concerns about political, regional, and economic biases were also raised upon reflection of the various types of rankers encountered in everyday life (e.g., WS3: \textit{“Avoid the biases that may arise from several factors like economic and industrial factors and political factors”}). Transparency and control were similarly foregrounded: participants expressed a desire to understand, customise (e.g., WS2: \textit{“Ability to customise and understand the 'ranking' parameters”)}, and filter out unverified results (e.g, WS3: \textit{“Remove unverified information”}).

These early articulations were important, as they shaped how participants evaluated rankings throughout the workshop. As such, they entered with strong normative expectations about what search systems ought to provide.

\subsection{Epistemic Authority of Neural Rankings and Undetected Manipulation}

Participants’ trust in the search system developed progressively across tasks, shaped by contrasts between lexical and neural ranking models. Early interactions with BM25 established a baseline perception of the system as limited. Participants frequently described BM25 as overly literal or dependent on keyword matching (e.g., WS3: \textit{“Only the literal meaning of `famous' is considered. I think in the prompt `famous doctor' there needs to be more importance on the [word] `doctor'”}; WS2: \textit{“But I don't expect tech to `know' exactly what I meant”}). While BM25 was often seen as failing to capture intended meaning, these failures were largely interpreted as mechanical limitations rather than as deceptive or harmful behaviour.

In contrast, when participants interacted with MonoT5, perceptions shifted markedly. The neural model’s apparent semantic coherence and fluent output led participants to treat its rankings as more credible and authoritative. Participants expressed satisfaction with the model’s interpretation of semantic intent (e.g., WS1: \textit{“This model is more suitable, as it provides a list of mostly people”}), and critiques focused less on relevance failures and more on systemic patterns within the results. Concerns centred on representational bias, such as homogeneity across gender, geography, or professional background, rather than on the correctness of the ranking in relation to the query (e.g., WS1: \textit{“Need to know the metric they use to rank/describe fame”}; WS2: \textit{“All dead? All men? Mostly, but not all European. Mostly, but not all medical”}). Importantly, these observations did not lead participants to question the legitimacy of the ranking mechanism itself. Instead, MonoT5 was largely perceived as working as intended, even when participants identified normatively problematic distributions or omissions within the output that could result in a biased information exposure.

This perceived legitimacy persisted into Task 4, where MonoT5 was subjected to deliberate adversarial manipulation. When presented with the query “Tylenol”, participants initially accepted the manipulated ranking as genuine (e.g., WS3: \textit{“These results seem quite relevant – good mix between definition of Tylenol and things related”}). Only when prompted multiple times by the workshop facilitators to examine individual results more closely did scepticism emerge regarding the anomalously ranked biography (e.g., WS1: \textit{“Provide reasons why that article was ranked in that position”}; WS2: \textit{“Helen Keaney – her snippet reads strangely, mentions Tylenol multiple times and even twice in the same list of brands advertised. Potentially hallucinated result?”}; WS3: \textit{“Stunt to attract advertising committees to being noticed (i.e., stunt by Helen Keaney)”}). Broader concerns about overreliance on neural systems surfaced primarily after the manipulation was revealed, at which point participants reflected on misplaced trust and system vulnerability (e.g., WS2: \textit{“I thought AI would be wiser. But we're expecting it to think like we do which is not fair. Anthropomorphic”}; WS3: \textit{“AI ranking models are more susceptible to manipulation than people might realise. An article that has obviously been manipulated for SEO came up 5th, which was problematic in this case. Unethical and not factual”}).

Together, these responses illustrate what can be described as a convincing wrongness, in which neural rankings appear coherent and convincing enough that distortions or manipulations remain undetected unless users are explicitly prompted to scrutinise them.

\begin{figure*}
  \centering
  \includegraphics[width=1\linewidth]{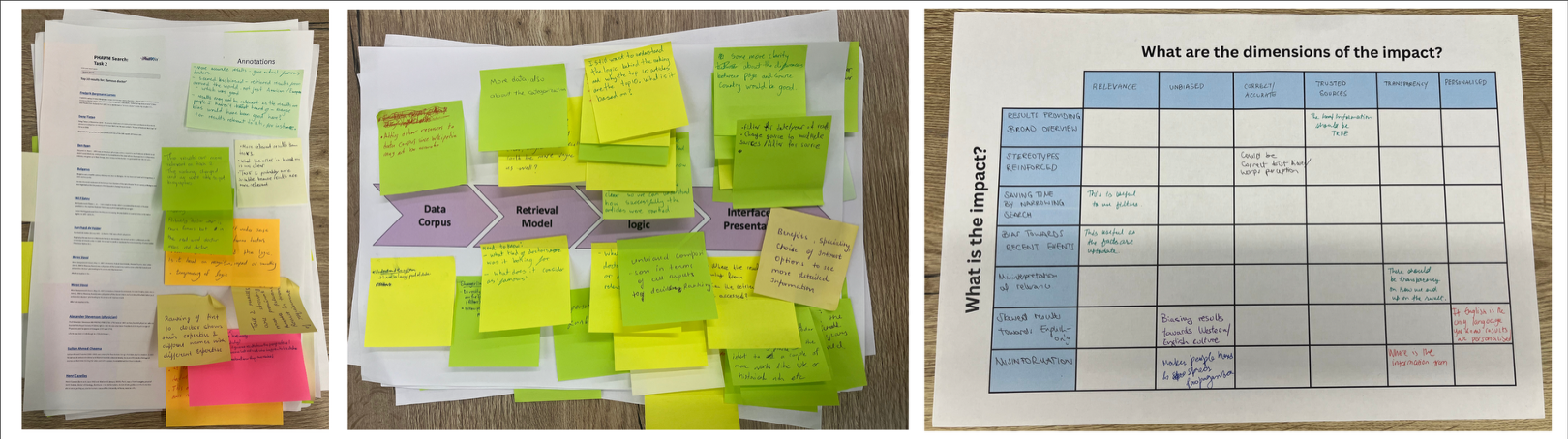}
  \caption{Annotated interface screenshot (left), search pipeline (middle), and matrix of impacts x dimensions of impact (right).}
  \Description{Annotated (directly on paper or with post-its) interface screenshot consisting of search bar and top ten results (left), simplified search pipeline from \autoref{fig:PipelineBlank} (middle), and matrix of impacts and dimensions of impact (right).}\vspace{-1em}
  \label{fig:Pipeline}
\end{figure*}

\subsection{Transparency and Recourse as Coupled Accountability Needs}

Across workshops, participants consistently framed search auditing as an interactive and corrective practice rather than a passive inspection of ranked outputs. Rather than merely wanting to understand why certain results appeared, participants expressed a desire to influence, correct, or report on system behaviour. Requests for \textit{“a feature that could narrow the [search] term”} (WS1), \textit{“a filter option to optimise the search output and reduce ambiguity”} (WS2) or \textit{“ranking logic to be clearer, so we can understand how successfully the articles were ranked”} (WS3) were expressed across the workshop groups in the simplified pipeline annotation task (\autoref{fig:Pipeline}). In this framing, accountability was understood less as post-hoc explanation and more as the availability of meaningful recourse: the ability to respond to, contest, or reshape rankings perceived as misleading, biased, or disproportionate.

Importantly, participants’ desire for recourse was closely tied to their attempts to understand where perceived problems originated within the search pipeline. During reflection activities following each task with our search system, participants repeatedly asked questions about the data corpus (WS2: \textit{“Where the article is sourced from that it's ranking”}) and the ranking logic (WS3: \textit{“How the results are ranked? Based on what standards?”}). These questions can be understood as interpretive tools through which participants tried to gain insight into the inner workings of the system. When results appeared unexpected or problematic, participants sought to triangulate responsibility by reasoning about whether issues stemmed from the corpus, the ranking model’s logic, or the interface and presentation of results. As such, for participants, auditing relied on constructing an interpretive understanding of the infrastructure that produced rankings.

Participants’ transparency needs were particularly pronounced at the level of corpus provenance. As one participant with experience in education reflected (WS2P5): \textit{“I'm thinking even back when we started to use the internet with [students] and we would say actually look at the URL and look where it's come from. A famous example was about Nelson Mandela and it was all these awful things that it told them all about him and you looked at who actually put this together, and it was a Ku Klux Klan website.”} Thus, concerns about provenance and agenda-setting long predate contemporary AI systems, yet remain central to how users assess credibility and harm in search.

However, transparency alone was viewed as insufficient. While participants welcomed visibility into elements of the search pipeline, this visibility often raised further questions about bias, omission, or misalignment, which the interface did not allow them to act upon. This tension became particularly evident in Task 3, where additional transparency and control features were introduced. Although these features helped participants recognise patterns of bias, they were frequently described as operating too late or at the wrong level of the pipeline to meaningfully address underlying issues. As one participant noted, mechanisms such as filtering results felt like \textit{“adding tape to fix a problem that occurred much earlier”} (WS1P3). In this sense, transparency without effective recourse risked creating an illusion of accountability, in which users could diagnose problems without being able to contest or correct them.

\subsection{Collective Sense-Making of Impacts of Search}

Following interaction with the custom search interface, participants collectively mapped perceived positive and negative events associated with search systems. Using a matrix (\autoref{fig:Pipeline}) that combined the features participants had earlier identified as mattering to them, (e.g., relevance, bias, and trustworthiness) with potential events and consequences of search (e.g., misinformation, political polarisation, and centralisation of power), participants populated the table collaboratively through group discussion.

Participants articulated causal narratives linking properties of search systems (along the dimensions that mattered to them) to downstream social effects. For example, bias and lack of diversity were associated with confirmation bias, stereotype reinforcement, blind spots, misinformation, propaganda, and political polarisation, as W2P1 expressed: \textit{“it enforces bias”}, supported by W2P2 who added: \textit{“I always think like you have to know the sum of the whole before you know what the part represents. So, if you think of a pie chart, you have to know what 100 looks like to know if 20 is a lot or a little”}. Limited transparency prompted questions about information provenance and trust. For instance, WS1P2 expressed: \textit{“There needs to be some kind of an AI authenticity score where users should be told whether there is an AI behind this”}. Failures of relevance or accuracy were framed as imposing additional cognitive and temporal burden on users, as echoed by WS2P3: \textit{“Less accurate and less relevant also waste more time.”} Centralised control over AI tools was linked to power concentration and the omission of alternative or dissenting perspectives, wherein some contexts may be more susceptible to the imbalances in their information provision than others, as WS3P4 expressed: \textit{“there are two or three companies only, you know most AI-tools are centralised, so they can use them for their own bias”}, to which WS3P7 added that: \textit{“when you see these companies change their direction, for instance towards the right wing, it might happen more so in the English context, like Western politics, not in the local ones that don't impact these kind of companies.”} Furthermore, AI-optimised content creation was framed as enabling manipulation and unfair advantage, especially for those who weaponise infrastructural weaknesses in bad-faith or for personal gain.

These discussions reveal that participants approached auditing as an exercise in systemic reasoning, treating search as an infrastructural actor shaping access to information, and thus public knowledge and awareness, rather than as a neutral retrieval tool. Importantly, normative values such as fairness were discussed less as a formal property of rankings and more as an outcome of these broader dynamics: who benefits, who is excluded, and who is able to influence or game the system. Positive effects, such as exploration or semantic broadening, were acknowledged but typically framed as conditional and dependent on user intent or system design.

To synthesise the range of harms and accountability concerns surfaced across workshops, we derived a taxonomy of impacts, grouping participant-identified issues according to the level at which harms were articulated: epistemic, representational, infrastructural, and downstream social (\autoref{tab:taxonomy}). It captures how users reasoned about both immediate ranking effects and longer-term consequences. This synthesis highlights the breadth of impacts that became legible through participatory engagement, beyond those typically captured through output-focused evaluation.

\begin{table*}[t]
\centering
\caption{Taxonomy of impacts and accountability concerns surfaced through participatory auditing of ranked search results.}
\small
\begin{tabular}{p{3cm} p{10cm}}
\toprule
\textbf{Impact Category} & \textbf{User-articulated concerns} \\
\midrule
Epistemic & Misinformation; confirmation bias; reduced critical scrutiny; misplaced trust in error-prone rankings; misalignment between user intent and system interpretation of queries; epistemic burden from evaluating credibility \\ \hline
Representational & systemic patterns of visibility; omission (and absence of minority/alternative perspectives); homogeneity in ranked results \\ \hline
Infrastructural & Opaque corpus provenance; lack of recourse mechanisms; susceptibility to manipulation across pipeline components; concentration of power in ranking providers  \\ \hline

Downstream social & Political polarisation; reinforcement of stereotypes; exclusion of marginalised perspectives; agenda-setting effects; unfair advantage through AI-optimised content \\

\bottomrule
\end{tabular}
\label{tab:taxonomy}
\end{table*}

\section{Discussion}

In this paper, we make the case for participatory auditing as a complementary practice to expert-led audits. Our workshop findings revealed two distinct but related forms of insight: first, they surface user-perceived and -experienced forms of harm and accountability issues that are difficult to anticipate through technical evaluation alone (RQ1); second, they reveal challenges and limitations users encounter when auditing search systems (RQ2).

Furthermore, our findings demonstrate that end-users of search systems situate impacts and harms in their personal contexts and draw anecdotal parallels to similar experiences. By tying in their personal contexts in conjunction with expertise in diverse (non-technical) domains, they articulate experienced impacts and downstream effects that may have gone unnoticed in traditional top-down audit practices or technical evaluations. 

These preconceptions influenced the harms and accountability issues users surface when auditing, answering RQ1: participants foregrounded normative questions about who benefits, who is excluded, and who bears the costs of how items are ranked. After critical discussion about the search systems, participants wanted to have a better understanding of underlying data influencing the ranking, but this transparency also highlighted a power imbalance between the user and the search system provider that participants felt they could not overcome.

These findings align with prior work that positions participatory auditing as a means of broadening dominant framings of harm and accountability beyond expert-defined categories~\cite{birhaneAIAuditingBroken2024d, becerra_sandoval_rethinking_2025, DiCampliSanVito2026ParticipatoryAuditing}. Furthermore, what emerged was a taxonomy of user-surfaced harms in search. While Abercrombie et. al ~\cite{abercrombieCollaborativeHumanCentredTaxonomy2024b} and Shelby et al. ~\cite{shelby_sociotechnical_2023} provide general-purpose harm taxonomies derived from incident databases and literature surveys, \autoref{tab:taxonomy} differs in two respects: it is domain-specific to ranked search and IR, and it emerges from participatory, in-situ auditing rather than post-hoc incident analysis. This means the categories it surfaces, particularly infrastructural harms such as opaque corpus provenance and susceptibility to adversarial manipulation, reflect how users actively reason about and attribute harm during interaction, rather than how harms are categorised after the fact by external observers.

Crucially, what participatory auditing surfaced, is epistemic legitimacy as users experience it, as opposed to traditional IR metrics which optimise for click prediction and aggregate measures like exposure disparity across ranked results. Traditional IR metrics measure point-in-time, query-level performance and cannot capture how users experience repeated exposure to biased results over time, or how context shapes harm perception. For instance, while MonoT5 performed well on standard IR metrics for “famous doctor”, matching its well known effectiveness on IR test collections \cite{mckechnieBiObjectiveNegativeSampling2024, chariEffectsRegionalSpelling2023}, participatory auditing revealed results were skewed toward male, European biographies. Participants also identified that consistently encountering Western-centric sources across multiple queries constituted epistemic harm, a cumulative effect invisible in single-query relevance scores. This is not to suggest participatory auditing replaces technical evaluation; rather, it surfaces the normative dimension of what metrics like relevance or fairness are proxies for, revealing where those proxies fall short of capturing lived user experience.

At the same time, the workshops revealed important potential limits to users’ capacity to audit, answering RQ2, particularly in interactions with the neural ranking model MonoT5, once it had previously performed satisfactorily and earned user trust. In the adversarial manipulation task, this accumulated trust reduced critical scrutiny, as despite being in an explicit audit setting, participants largely accepted manipulated results unless explicitly prompted to examine them closely. This finding complicates assumptions that audit framing or context alone is sufficient to activate critical engagement. Instead, auditability appears shaped by judgements of trust, perceived system competence, and prior interaction history. Importantly, these limits were not revealed by participatory auditing alone, but emerged through observation of participants’ failures and blind spots during the workshops themselves.

Taken together, these findings show that participatory auditing plays a dual and reflexive role. First, it enables the discovery of harms and impacts that users themselves associate with search systems, including downstream social and epistemic effects that are rarely surfaced through output-based or expert-led audits~\cite{birhaneAIAuditingBroken2024d}. Second, it exposes the conditions under which auditability itself is enabled and those in which it breaks down. In our workshops, participants’ prior experiences with the seemingly competent MonoT5 shaped what they attended to and what they overlooked, leaving them susceptible to content-level manipulation that exploited trusted components of the pipeline. Beyond simply revealing accountability failures, participatory auditing also surfaces how users are conditioned to look for certain kinds of errors and harms while missing others.

\section{Limitations}
This study has limitations that should be considered when interpreting the findings. First, the workshops varied in size and participant numbers. While this variability is consistent with qualitative research that prioritises discussion-depth over statistical generalisability, it may have influenced the dynamics of group deliberation across sessions. Furthermore, we recognise that future work would be enriched by broader stakeholder inclusion. For this paper, we focused on end-users as the primary affected party as a starting point, consistent with participatory audit literature beginning with those experiencing direct impact \cite{birhaneAIAuditingBroken2024d}. As such, future work should include content creators, marginalised communities not represented in our sample, and platform workers.

Second, the workshops were conducted under time constraints and necessarily provided only a brief introduction to the concept of auditing search systems. As a result, participants engaged in an artificial auditing context, and their understanding of auditing likely varied. Rather than constituting a shortcoming of the study alone, this reflects a broader challenge for participatory auditing: critical engagement and audit literacy cannot be assumed and may need to be actively supported for participatory audits to succeed.

Third, the search queries were pre-selected by the researchers. While this enabled controlled comparison across ranking architectures and conditions, some queries may not have aligned with participants’ personal information needs or expertise. This may have limited their ability to assess relevance or quality in ways that reflect their everyday search behaviour.

Finally, the scope of this work focused on ranked search results rather than on emerging forms of search interaction, such as retrieval-augmented generation~\cite{LewisRAG} or conversational search interfaces~\cite{RadlinskiConvSearch}. As search systems increasingly integrate ranking with generative components, future work should examine auditability in these hybrid systems.

\section{Conclusions}

This paper examined what becomes visible when users are invited to audit ranked search systems through participatory engagement. Across three workshops, we showed that participatory auditing enables the discovery of impacts, harms, and accountability concerns that remain difficult to surface through output-focused or expert-led audits alone. Participants drew on lived experience, personal context, and domain knowledge to articulate impacts spanning epistemic, representational, infrastructural, and downstream social dimensions, situating search systems within broader questions of trust, power, and access to information.

Beyond impact discovery, this paper contributes an empirical account of the limits of participatory auditing itself. We show that ranking models perceived as interpretively adequate can suppress critical scrutiny by appearing to work as intended, even when they manipulated outputs. Through the adversarial injection task, we demonstrate how prior interaction history and accumulated trust shape what users attend to during audits, leading to misrecognition of harms. Importantly, these limits were not revealed by participants themselves, but were observed through participants’ failures, misinterpretations, and blind spots during the workshops themselves.

Taken together, we argue that participatory auditing should be understood as a complementary auditing mode: one that both broadens what counts as harm by foregrounding lived and situated experience of diverse individuals, and exposes the sociotechnical conditions under which fairness and accountability practices succeed or break down. As regulatory frameworks such as the EU AI Act and the DSA increasingly emphasise ongoing risk assessment, transparency, and user-facing accountability for AI systems, participatory approaches offer a critical means of grounding such requirements in real-world use and contexts after deployment.

\begin{acks}
This work was supported by the Engineering and Physical Sciences Research Council [grant number EP/Y009800/1], through funding from Responsible AI UK (KP0011). We wish to thank Mahdi Dehghan and Ritajit Dey for their support in facilitating the workshops and James Nurdin and Zeyan Liang for participating in the piloting of the workshop activities.
\end{acks}

\bibliography{reference2}
\bibliographystyle{acm}

\appendix
\section{Appendix}
\subsection{Workshop Protocol}

\begin{table}[h!]
    \centering
    \caption{Session Overview}
    \begin{tabular}{|p{2.7cm}|p{4.8cm}|}
    \hline
       \textbf{Participant Count} & WS1: 4; WS2: 9; WS3: 8 \\ \hline
        \textbf{Format} & In-person group workshop with individual device-based search tasks\\ \hline
        \textbf{Devices} & Participants' own smartphones or researcher provided tablets \\ \hline
        \textbf{Recording} & Audio and video recordings, post-its \\ \hline
        \textbf{Segment 1} & Introductory Auditing Activities \\ \hline
        \textbf{Segment 2} & Standard Identification \\ \hline
        \textbf{Segment 3} & Tasks 1-4 with Reflection Activities\\ \hline
        \textbf{Segment 4} & Collective Impact Mapping\\ \hline
    \end{tabular}
    \label{tab:appendix}
\end{table}

\subsection{Segment 1: Introductory Auditing Activities}
\textbf{Duration: 20min}

\subsubsection{Introduction to Auditing}
Facilitators presents a brief introduction to algorithmic auditing using slides, covering: what an audit is, common audit types (financial, operational, privacy, compliance, AI), and the purpose of participatory auditing. Participants are introduced to the PHAWM project and the workshop objectives.

\subsubsection{Policy Audit Example: DSA Article 27(1)}
Participants collectively audit TikTok against EU Digital Services Act (DSA) Article 27(1), which requires platforms to disclose recommender system parameters in plain and intelligible language and provide options to modify or influence those parameters. The facilitators walk through TikTok’s support documentation and terms, asking participants to assess whether TikTok meets each sub-requirement. This activity establishes shared auditing vocabulary and demonstrates the difference between adequate and inadequate compliance. Key discussion points:
\begin{itemize}
    \item Does TikTok explain its recommender parameters in plain language? (First part of Article 27(1))
    \item Does TikTok provide user-facing controls to modify those parameters? (Second part of Article 27(1))
    \item What counts as a meaningful “option” versus passive user interaction?
\end{itemize}

\subsubsection{Live Wikipedia Search}
Before using the custom search interface, participants view a live Wikipedia search\footnote{https://en.wikipedia.org/w/index.php?search} for the query “chemist”. Participants are asked to scroll through the first 50–100 results and then discuss:
\begin{itemize}
    \item Do you notice anything about the results?
    \item Do you accept that these are the most relevant results?
    \item Is anything missing from the results?
    \item What are the consequences of this type of ranking?
\end{itemize}

This step draws attention to ambiguity (chemist as scientist vs. drugstore), gender and ethnic diversity gaps in biographical results, and position bias. It primes participants to think critically about search before interacting with the custom interface.

\subsection{Segment 2: Standard Identification}
\textbf{Duration: 15min}
\hfill \break
Before participants interact with the custom search interface, they individually complete a standards identification activity. Each participant receives a set of blank post-it notes and given the following prompts:

\begin{itemize}
    \item What matters to you in ranked search results?
    \item Think about: fairness, unbiased ranking, accuracy, transparency, relevance, or anything else that comes to mind.
\end{itemize}

Participants place post-its on a shared whiteboard. The facilitators briefly reviews the responses before moving to Task 1. These post-its serve as baseline normative expectations that are later used to populate the horizontal axis of the final impact matrix (Segment~4). In auditing terms, these are the standards that they wish to audit the search system against.

\subsection{Segment 3: Tasks 1-4}
\textbf{Duration: 60min}
\subsubsection{Task 1: Baseline Lexical Ranker (BM25)}
\hfill \break
\textbf{Query: “famous doctor”}

The purpose of the task is to establish a baseline for participants’ expectations around relevance, system capability, and keyword-driven ranking. BM25 produces intentionally poor results for this query due to its literal keyword matching, which cannot handle the semantic interpretation required by the adjectival qualifier “famous”.

Participant instructions are as follows:

\begin{itemize}
    \item Navigate to the custom Search interface (Task 1 version).
    \item Enter the query “famous doctor” and review the top 10 results.
    \item Read through the result titles and snippets.
    \item Annotate a printed screenshot of the results page with observations.
\end{itemize}

The reflection prompts are as follows:

\begin{itemize}
    \item Which articles are ranking well? Why do you think they ranked well?
    \item How relevant to your query do you think the top 10 results are?
    \item Do you accept that these top results are the most relevant? Why or why not?
    \item Do you think anything is missing from the results?
    \item What consequences (positive or negative) could occur as a result of how this ranking works?
    \item What information or data would you need to know about how these articles are ranked in order to have an informed opinion?
\end{itemize}

\subsubsection{Task 2: Neural Semantic Reranker (MonoT5)}
\hfill \break
\textbf{Query: “famous doctor”}

The purpose of the task is to compare a neural semantic reranker with the lexical baseline. MonoT5 in comparison produces more coherent, biography-focused results. Participants compare the two models and reflect on what additional transparency and control they would need to audit such a system.

Participant instructions are as follows:

\begin{itemize}
    \item Navigate to the Task 2 version of the custom Search interface.
    \item Enter the same query “famous doctor” using the new ranking model.
    \item Review the top 10 results and compare with Task 1.
    \item Annotate a printed screenshot of the Task 2 results.
\end{itemize}

The reflection prompts are as follows:

\begin{itemize}
    \item Did you perceive a difference between the two ranking models?
    \item How did the rankings change between Task 1 and Task 2?
    \item Which model is more suitable to your query, and why?
    \item How relevant do you think the top 10 results are this time?
\end{itemize}

Participants are then shown simplified pipeline diagram of search. In relation of the pipeline diagram, they receive following instructions: 

\begin{itemize}
    \item Reflect on where would you need more information to understand the system? (Yellow post-its)
    \item What changes or improvements would you want to see? (Green post-its)
    \item Think about which component of the pipeline (data corpus, retrieval model, ranking logic, interface) each need or improvement applies to. Stick your post-its on the respective components they apply to.
\end{itemize}

\subsubsection{Task 3: Transparency and User Control}
\hfill \break
\textbf{Query: “famous doctor”}

The purpose of the task is to explore how varying levels of interface transparency and user control affect participants’ ability to audit and contest search results. The Task 3 interface adds: model switching (BM25 labelled as "Generic Keyword Search" and MonoT5 labelled as "AI-Powered Ranking"), article metadata display (page country, source country, page views, language availability), metadata-based filters, and an auto-fairness re-ranking function.

Participant instructions are as follows:

\begin{itemize}
    \item Before entering a query, inspect the new Task 3 interface carefully.
    \item Reflect on what outcomes you would expect from each available control. Note your expected outcomes on a post-it (Expected outcomes column of annotation poster).
    \item Enter the query “famous doctor” and explore the interface controls.
    \item After using the controls, compare the actual outcomes to your expectations. Note actual outcomes (Actual outcomes column of annotation poster).
    \item Reflect on what outcomes you would ideally want in a search system.
\end{itemize}

The reflection prompts are as follows:

\begin{itemize}
    \item (Before using the controls) What outcomes do you expect from each available control?
    \item (After using the controls) What was the actual outcome? Did it match your expectation?
    \item What outcomes would you actually want in an ideal system?
    \item Would you want other insights beyond those shown? What additional transparency would help?
    \item Would you want other controls?
    \item Would you want to re-rank results based on other metrics or criteria?
    \item Returning to the pipeline diagram: how would you change the system to better understand it and express your preferences? [Annotate the model]
\end{itemize}

\subsubsection{Task 4: Adversarial Manipulation}
\hfill \break
\textbf{Query: “Tylenol”}

The purpose of the task is to assess whether participants can detect adversarial manipulation in a neural ranking. One Wikipedia biography (“Helen Keaney”, normally ranked 17th) has been modified via keyword injection: the terms “relevant”, “true”, and “Tylenol” were added to its title and snippet to exploit a known susceptibility in MonoT5’s inference-time prompt. Only the top 5 results are shown. Participants are not told in advance that manipulation has occurred.

Participant instructions are as follows:

\begin{itemize}
    \item Enter the query “Tylenol” and carefully read the snippets of all top 5 results.
    \item Annotate the results screenshot provided with your observations (before the facilitator reveals the manipulation).
\end{itemize}

The reflection prompts are as follows:

\begin{itemize}
    \item What do you notice about the results? Read all 5 snippets carefully.
    \item Do the results seem as expected for a query about a pharmaceutical product?
    \item Is any result surprising or out of place? If so, why?
    \item (After manipulation revealed) What does this tell you about the weaknesses and susceptibilities of the AI ranker?
    \item Would you have expected a semantic ranker (trained to understand meaning rather than keywords) to be more robust against this kind of linguistic manipulation?
    \item Has this task challenged your assumptions about AI ranking models? If so, how?
    \item What recourse mechanisms would you want if you encountered a manipulated result in a real search system?
\end{itemize}

\subsection{Collective Impact Mapping}
\textbf{Duration: 25min}
\subsubsection{Cause-Effect Mapping}
Participants are asked to identify consequences (positive and negative) that search systems can have on themselves personally, their communities, and society. They write causes and effects on post-its and place them in a two-column Cause~|~Effect sheet. One example is provided by the facilitator to prompt discussion: Semantic AI rankers can be influenced through keyword injections (cause), as a result a biased or manipulated article ranks highly, influencing user perception (effect). Participants are encouraged to connect effects back to pipeline components where possible.

Discussion prompts:

\begin{itemize}
    \item What consequences (positive or negative) can search have on you personally, your friends, family, wider community, and on society more generally?
    \item Think of both positive effects (e.g., discovering useful or trustworthy information) and negative effects (e.g., reinforcing bias, amplifying misinformation).
    \item How do these effects change when search becomes or is perceived as more “intelligent”?
\end{itemize}

\subsubsection{Impact Matrix}
The facilitator introduces a shared impact matrix. The vertical axis (“What is the impact?”) is pre-populated with the effects participants identified in Cause-Effect Mapping exercise. The horizontal axis (“What are the dimensions of the impact?”) is populated using the dimensions participants identified as mattering to them during the baseline elicitation in Segment 2 (e.g., relevance, bias, accuracy, transparency). Participants collectively populate the matrix cells through facilitated group discussion, mapping search properties to perceived events and downstream consequences. This activity is designed to surface how participants connect specific consequences related to ranking to broader social and epistemic effects.

\section{Endmatter}

\subsection{Generative AI Usage Statement}
The authors used institutionally approved AI-tools, Microsoft Copilot (integrated with institutional enterprise account) and Grammarly plug-in on Overleaf, solely for grammar, spelling, and fluency at the sentence level. These tools were not used to generate text, arguments, or content for this paper. All wording, analysis, and interpretations were authored by the authors, who take full responsibility for the integrity of the work.

\subsection{Author Contributions}
\textbf{Anna Marie Rezk}: Conceptualisation, Methodology, Software, Formal analysis, Investigation, Writing - Original Draft. \textbf{Patrizia Di Campli San Vito}: Investigation, Writing - Original Draft. \textbf{Ayah Soufan}: Investigation, Writing - Original Draft. \textbf{Graham McDonald}: Software, Resources, Writing - Review \& Editing. \textbf{Craig Macdonald}: Software, Resources, Writing - Review \& Editing. \textbf{Iadh Ounis}: Resources, Supervision.

\subsection{Competing Interests}
All authors were at the time of conducting the research, working together on the Participatory Harms Auditing Workbenches and Methodologies (PHAWM) project. Funding details are in acknowledgements.

\subsection{Positionality Statement}
The lead author is a researcher in human-computer interaction with a background in media studies, recommender systems, and responsible AI, which shapes an orientation toward accountability, usability, and fair access to information in algorithmic systems. This background informed the framing of search as an infrastructural technology with normative consequences. The methodological approach of participatory auditing was further shaped by the authors' roles as researchers on the PHAWM project, a funded initiative specifically focused on developing participatory auditing workbenches and methodologies for AI systems. The co-authors bring expertise in participatory AI auditing, human-computer interaction, and information retrieval. As researchers embedded in academic institutions, we recognise that our participant pool was drawn from a university campus, which limits the diversity of perspectives captured and reflects our institutional context. We have attempted to mitigate this by enrolling participants across varied professional and educational backgrounds, but acknowledge that future work should engage more diverse communities.

\subsection{Ethical Considerations Statement}

This study involved human participants recruited through posters and mailing lists at the University of Glasgow. All participants provided written informed consent prior to taking part, were informed of their right to withdraw at any time without consequence, and received £40 GBP compensation for their time. The workshop methodology, including consent procedures, data collection, and compensation, was approved by our university's Ethics Committee prior to data collection. Sessions were audio- and video-recorded with participants' explicit consent; recordings were stored securely on institutional servers accessible only to the research team and will be deleted following the project's data retention period. No sensitive personal data beyond demographic information was collected. Participant quotations are reported using workshop and participant identifiers (e.g., WS1P1) rather than names to protect anonymity. The custom search interface used a publicly available corpus (Wikipedia, via the TREC Fair Ranking Track) and did not collect or store participants' search queries. The adversarial manipulation task involved deliberate deception in that participants were not informed in advance that one result had been manipulated. Importantly, the manipulation was conducted solely within a controlled research prototype deployed locally for the purposes of this study and was not applied to any live, publicly accessible system. We judged this necessary to study authentic detection behaviour, and the manipulation was fully disclosed to all participants at the end of the task. No participant expressed distress or concern about this deception upon disclosure. We do not anticipate adverse impacts from publication of this work; the adversarial manipulation technique exploited is already documented in the IR literature \cite{parry_analyzing_2024}, and its disclosure here is intended to raise awareness rather than enable harm.

\end{document}